\documentclass{article}
\usepackage[T1]{fontenc} 
\usepackage[utf8]{inputenc} 
\usepackage{ismir,amsmath,cite,url}
\usepackage{graphicx}
\usepackage{color}

\usepackage{arydshln}
\usepackage{dblfloatfix}

\usepackage{amsfonts}

\title{Vector-Quantized Timbre Representation}

\threeauthors
  {Adrien Bitton\textsuperscript{1,2}\thanks{\textsuperscript{1} Institut de Recherche et Coordination Acoustique Musique, CNRS UMR9912 (IRCAM), Paris, France.}} {{\tt bitton@ircam.fr}}
  {Philippe Esling\textsuperscript{1}} {{\tt esling@ircam.fr}}
  {Tatsuya Harada\textsuperscript{2}\thanks{\textsuperscript{2} Machine Intelligence Laboratory (MIL), The University of Tokyo, Japan.}} {{\tt harada@mi.t.u-tokyo.ac.jp}}

\begin{document}

\maketitle

\begin{abstract}

Timbre is a set of perceptual attributes that identifies different types of sound sources. Although its definition is usually elusive, it can be seen from a signal processing viewpoint as all the spectral features that are perceived independently from pitch and loudness. Some works have studied high-level timbre synthesis by analyzing the feature relationships of different instruments, but acoustic properties remain entangled and generation bound to individual sounds. This paper targets a more flexible synthesis of an individual timbre by learning an approximate decomposition of its spectral properties with a set of generative features. We introduce an auto-encoder with a discrete latent space that is disentangled from loudness in order to learn a quantized representation of a given timbre distribution. Timbre transfer can be performed by encoding any variable-length input signals into the quantized latent features that are decoded according to the learned timbre. We detail results for translating audio between orchestral instruments and singing voice, as well as transfers from vocal imitations to instruments as an intuitive modality to drive sound synthesis. Furthermore, we can map the discrete latent space to acoustic descriptors and directly perform descriptor-based synthesis.

\end{abstract}

\section{Introduction}\label{sec:intro}

\textit{Timbre} is a central element in musical expression and sound perception \cite{timbre_mcadams}, which can be seen as a set of spectral properties that allows us to distinguish instruments played at the same pitch and velocity. Synthesis of musical timbre has been studied by analyzing the feature relationships between instruments. A disentangled representation of pitch and timbre was proposed in \cite{ismirtimbre} which allows to generate musical notes with instrument control. Perceptual timbre relationships were explicitly modeled in \cite{dafxtimbre}, and latent timbre synthesis could be iteratively mapped to target acoustic variations. However, both techniques are not evaluated in the signal domain and acoustic properties remain entangled. A timbre-invariant representation of variable-length waveforms is learned in \cite{musictranslation} to perform unsupervised translation of an instrument performance into another, which we refer to as \textit{timbre transfer}. However, such representation is not interpretable and does not offer any controls besides selecting a target instrument class.

This paper introduces a generative model training on an individual timbre domain that allows variable-length timbre transfer of diverse audio sources and sound synthesis with direct acoustic descriptor control. This auto-encoder with a discrete latent space that is disentangled from loudness learns the feature quantization of a given timbre distribution. Latent features are decoded into short-term spectral coefficients of a filter applied to overlapping frames of a noise excitation. This subtractive synthesis technique does not constrain the types and lengths of signals that can be processed. We perform timbre transfer by encoding any input signals into this discrete representation. The matched series of latent features is inverted into a signal which corresponds to the trained timbre domain. Since the model has learned an approximate decomposition of a timbre into a set of short-term spectral features, we can individually decode each latent vector and compute the corresponding acoustic properties. It provides a direct mapping for \textit{descriptor-based} synthesis. A descriptor target can be matched with a series of latent features and decoded into a signal with the desired auditory property.

Our timbre transfer experiments apply to orchestral instruments and singing voice. We pretrain an instrument classifier and evaluate transfer with the predicted accuracy of a model at translating all other instruments into the trained timbre domain. And we measure the distances between the input and output fundamental frequency and loudness. These distances amount to the error of a model at preserving the source pitch and loudness independently from transforming the timbre. We also perform timbre transfer from vocal imitations to instruments as an example of voice driven synthesis. Whereas many sound ideas are hardly described with musical parameters, which require an expert knowledge, human voice control can be an intuitive medium \cite{imit_sound}. For instance, mimicking some moods, objects or actions that are translated into musical sounds.

\section{State of the art}\label{sec:sota}

\subsection{Generative Modeling}\label{subsec:gen_model}
Generative neural networks aim to model a given set of observations $\mathbf{x}\in\mathbb{R}^{d_{x}}$ in order to consistently produce novel samples $\tilde{\mathbf{x}}$. To this extent, we introduce latent variables $\mathbf{z}\in\mathbb{R}^{d_{z}}$ defined in a lower-dimensional space ($d_{z}<d_{x}$). These latent features form a simpler representation from which the data can be generated. An unsupervised approach to learn these variables is the \textit{auto-encoder}. A deterministic encoder maps observations to latent codes $\mathbf{z}=E_{\phi}(\mathbf{x})$ that are fed to the decoder which in turn reconstructs the input $\tilde{\mathbf{x}}=D_{\theta}(\mathbf{z})$. Their parameters jointly optimize some reconstruction loss
\begin{equation}
\underset{\phi,\theta}{\text{argmin}}\;\mathcal{L}_{rec.}\left(\mathbf{x},D_{\theta}\left(E_{\phi}\left(\mathbf{x}\right)\right)\right).
\label{eq:ae_rec}
\end{equation}
As this approach explicitly performs dimensionality reduction, these latent variables can extract the most salient features in the dataset. Hence, they also facilitate the generation over high-dimensional distributions. However, in this deterministic auto-encoder setting there is no guarantee that latent inference on unseen data produces meaningful codes for the decoder. In other words, these latent projections are usually scattered apart from those of the training observations, and the decoder may fail at reconstructing anything consistent besides its training domain.

Regularized auto-encoders tackle this problem by introducing constraints over the distribution of latent codes and generation mechanism. To do so, the Variational Auto-Encoder (VAE)\cite{VAE} sets a probabilistic framework by optimizing a variational approximation of the encoder distribution $q_{\phi}(\mathbf{z}|\mathbf{x})$ given a continuous prior $p_{\theta}(\mathbf{z})$ over latent variables. The model is trained with a Kullback-Leibler (KL) divergence regularizer added to a reconstruction cost
\begin{equation}
\resizebox{.89\hsize}{!}{$\mathcal{L}_{\theta,\phi}=\underbrace{-\mathbb{E}_{q_\phi(\mathbf{z})}\big[ \log{ p_\theta (\mathbf{x|z}) } \big]}_{\text{reconstruction}}+\underbrace{\mathcal{D}_{KL} \big[ q_\phi(\mathbf{z|x}) \parallel p_\theta(\mathbf{z}) \big]}_{\text{regularization}}$}.
\label{eq:vae_obj}
\end{equation}
VAEs provide several desirable features such as their interpolation quality, generalization power from small datasets, and the ease for both high-level visualization and sampling. However, they tend to produce less detailed low-level features (blurriness effect), and the regularization can degenerate into an uninformative latent representation (a phenomenon known as \textit{posterior collapse} \cite{postcollapse}).

\begin{figure}[ht]
\centering
\includegraphics[width=1.\columnwidth]{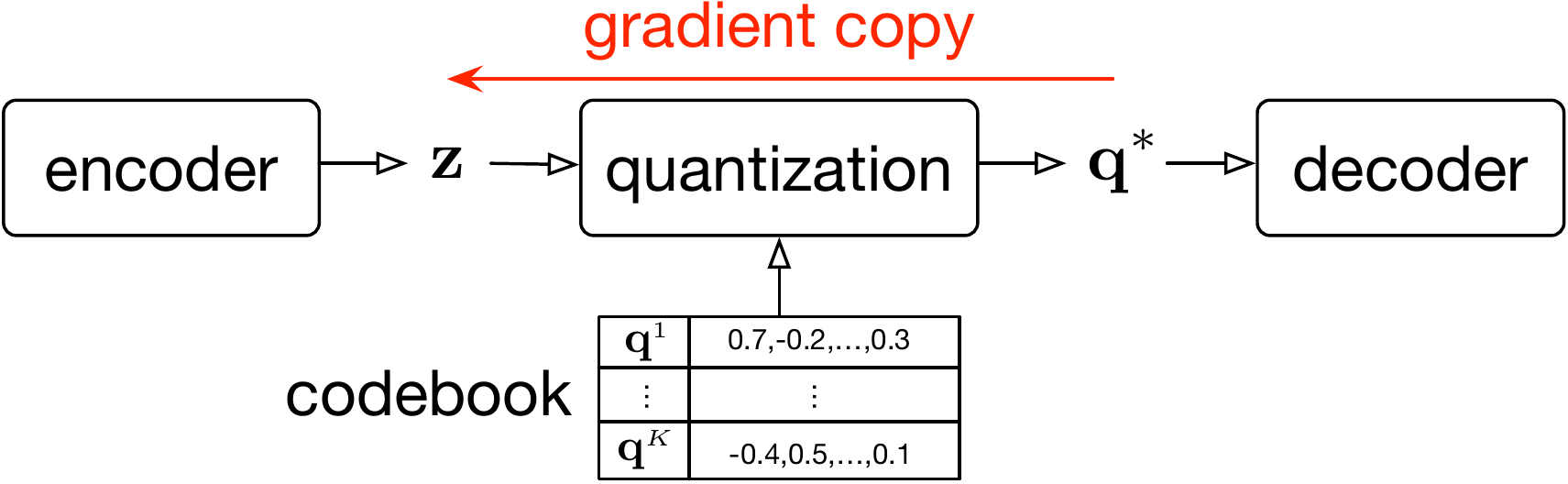}
\caption{Overview of a Vector-Quantization (VQ) layer.\label{fig:VQVAE}}
\end{figure}

The Vector-Quantized VAE (VQ-VAE)\cite{vqvae} addresses these issues by learning a \textit{discrete} latent representation, defined as a \textit{codebook} with a fixed number of latent vectors $\{\mathbf{q}^{_1},\ldots,\mathbf{q}^{_K}\}$. Hence, the output $\mathbf{z}$ of the deterministic encoder is matched to its nearest embedding code $\mathbf{q}^*$
\begin{equation}
\mathbf{q}^* = \underset{j\in[1,K]}{\text{argmin}} \left\|\mathbf{z}-\mathbf{q}^{_j}\right\|_2
\label{eq:near_code}
\end{equation}
which is passed to the decoder, so that it optimizes generation solely using the current codebook state. In addition to the latent dimensionality reduction, the amount of information compression is set by the size $K$ of the discrete embedding. Assuming a uniform prior distribution over the embedding, the amount of information encoded in the representation corresponds to a constant KL divergence of $\log(K)$. Since that hyperparameter is not optimized, the VQ-VAE alleviates posterior collapse.
The representation is optimized with a \textit{codebook update} loss which matches the selected code to the encoder features
\begin{equation}
\mathcal{L}_{codebook}=\left\|sg(\mathbf{z})-\mathbf{q}^*\right\|^2_2
\label{eq:update_code}
\end{equation}
where $sg$ denotes a \textit{stop gradient} operation, bypassing the variable in the back-propagation. Symmetrically, the \textit{encoder commitment} to the selected code is applied as a loss
\begin{equation}
\mathcal{L}_{commit}=\left\|\mathbf{z}-sg(\mathbf{q}^*)\right\|^2_2
\label{eq:commit}
\end{equation}
in order to bound its outputs and stabilize the training. The complete objective with commitment cost $\beta$ is then
\begin{equation}
\mathcal{L}_{\text{VQ-VAE}} = \mathcal{L}_{rec.}(\mathbf{x},\tilde{\mathbf{x}}) + \mathcal{L}_{codebook} + \beta*\mathcal{L}_{commit}.
\label{eq:vqvae_obj}
\end{equation}
Because of the argmin operator (nearest-neighbor selection), Eqn \eqref{eq:near_code} is not differentiable and the encoder cannot be directly optimized. However, as shown in \figref{fig:VQVAE}, this issue is circumvented by simply copying the gradient from  $\mathbf{q^*}$ to $\mathbf{z}$ (straight-through approximation) and back-propagating this information in the encoder unaltered with respect to the quantization output. The VQ-VAE achieves sharper reconstructions than those of the probabilistic VAE, and its discrete latent representation was successfully applied to speech for unsupervised acoustic unit discovery\cite{vqspeech}. In this paper, it was shown that the quantized codebook could extract high-level interpretable audio features that strongly correlate to phonemes, with applications for voice conversion. Inference is performed by quantizing every continuous encoder outputs with the learned latent codebook. Consequently, the decoder is bound to reconstruct the input given this discrete latent space, whose degrees of freedom can be adjusted with the codebook size $K$. This reconstruction with latent quantization may be seen as a transfer when matching any out-of-domain inputs with a set of features learned from a given dataset.

\subsection{Raw Waveform Modeling}\label{subsec:raw_model}
The first methods for neural waveform synthesis have relied on auto-regressive sample predictions, as in the reference WaveNet model \cite{wavenet}. It achieves high-fidelity sound synthesis, at the cost of a heavy architecture that is inherently slow to train and sample from. In more recent developments, waveform models have exploited digital signal processing knowledge, providing efficient solutions that achieve competitive audio quality. It results in more interpretable and lighter architectures which consequently require less data to train on. A sinusoids plus stochastic decomposition\cite{sms} is first used in the Neural Source Filter (NSF) \cite{nsf} model. It generates speech from acoustic features and the estimated fundamental frequency $f_0$, that are used as a conditioning information for the synthesis modules. These are a sinusoidal source controlled by the $f_0$, a Gaussian noise source and two separate temporal filters to process each of them. The generated signals are adaptively mixed in order to render both voiced (periodic) and unvoiced (aperiodic) speech components. More specific to musical sound synthesis, the Differentiable Digital Signal Processing (DDSP)\cite{ddsp} model implements a similar decomposition with an additive synthesizer conditioned with $f_0$ summed with a subtractive noise synthesizer, both controlled by the decoder. It predicts the harmonic amplitudes and the frequency domain coefficients $\mathbf{H}_t$ to generate the filtered audio $\mathbf{y}_t$ from non-overlapping frames of noise $\mathbf{x}_t$
\begin{equation}
\mathbf{y}_t=\text{DFT}^{-1}(\mathbf{H}_t.\text{DFT}(\mathbf{x}_t))
\label{eq:DDSP}
\end{equation}
with $\text{DFT}$ the Discrete Fourier Transform and $\text{DFT}^{-1}$ its inverse. This model offers promising results and an interesting modularity that disentangles harmonic, stochastic as well as reverberation features. However, it is mainly tailored for harmonic sounds and does not allow end-to-end training as it relies on an external $f_0$ estimator.

The two aforementioned models train on a multi-scale Short-Term Fourier Transform (STFT) reconstruction objective, that is computed for several resolutions. The distance between spectrogram magnitudes is an efficient criterion for optimizing waveform reconstruction as it provides a structured time-frequency representation. However, since the phase is discarded, it may fail at evaluating certain acoustic errors. Based on human ratings to evaluate just-noticeable distortions, a differentiable audio metric is proposed \cite{perceploss} in order to assess artifacts at the threshold of perception. Listeners were asked whether pairs of audio were exactly similar, with one element being applied varying strengths of additive noises, reverberation or equalization. This dataset provides pairs of waveforms along with binary ratings, on which a convolutional neural network learns a differentiable loss. A deep feature distance $d$ is trained by forwarding each audio $\mathbf{x}$ (clean) and $\tilde{\mathbf{x}}$ (altered) into the network. Considering $L$ layers and $F_l \in \mathbb{R}^{T_l*C_l}$ the $l$-th convolution activations, it computes
\begin{equation}
d(\mathbf{x},\tilde{\mathbf{x}})=\sum_{l=1}^{L}{\frac{1}{T_l}
\left\|\mathbf{w}_l\odot(F_l(\mathbf{x})-F_l(\tilde{\mathbf{x}}))\right\|_1}
\label{eq:percep_dist}
\end{equation}
with $\mathbf{w}_l$ a learnable weight for each of the $C_l$ channels of width $T_l$. Given this deep feature distance, a low-capacity classifier is trained to infer human ratings of noticeable dissimilarity. In this setting, the network must efficiently model such just-noticeable differences in order to allow an accurate prediction. Once trained, this distance can be used as a differentiable audio loss. It was shown to improve the performance of speech enhancement systems and may be added as an additional reconstruction objective.

\subsection{Musical Timbre Transfer}\label{subsec:timbre_transfer}
The task of musical timbre transfer is to convert the identity of one sound into another, e.g. two instruments, while preserving independent features such as pitch and loudness. The model in \cite{ismirtimbre} learns a representation that disentangles these features inside instrument sounds. It offers interesting visualizations and generative controls. However, it is restricted to processing individual notes of limited duration from spectrogram magnitudes. As a result, synthesis occurs with an inversion latency and is not evaluated in the signal domain.

In this work we focus instead on unsupervised transfer for variable-length waveforms, such as recorded music performances. The Universal Music Translation Network \cite{musictranslation} proposes an architecture for multi-domain transfers, using a shared encoder paired with domain-specific decoders. The generalization of the learned representation to many domains is achieved with a latent confusion objective. It uses an adversarial classifier to enforce the domain-invariance of latent codes. The task is solved in the waveform domain by relying on multiple WaveNet models. For that reason, both training and synthesis are slow and computationally very intensive. Although it allows high-quality auto-encoding with domain selection, its latent representation does not offer more generative controls. On the other hand, more expressive and light-weight synthesis models can perform timbre manipulations with additional constraints. The DDSP model was applied to single domain transfer with independent control over pitch and loudness, but with limitations of its amortized inference.

\section{Vector-Quantized Model for Timbre}\label{sec:NSUB}

In this paper, we introduce a waveform auto-encoder for learning a discrete representation of an individual timbre that can be used for sound transfer and descriptor-based synthesis. We merge the VQ-VAE approach with a decoder that performs subtractive noise filtering with a disentangled gain prediction. As the model is unsupervised, it can train on diverse music performance recordings and can as well process non-musical audio such as vocal imitations. The resulting latent representation decomposes spectral timbre properties, while being invariant to loudness. The model performs timbre transfer by encoding any audio sources into the loudness-invariant feature quantization which is inverted to the learned timbre. The discrete latent space can be mapped to acoustic descriptors. It allows us to order series of latent features according to a descriptor target and offers meaningful synthesis controls.

\begin{figure*}[ht]
\centering
\includegraphics[width=1.6\columnwidth]{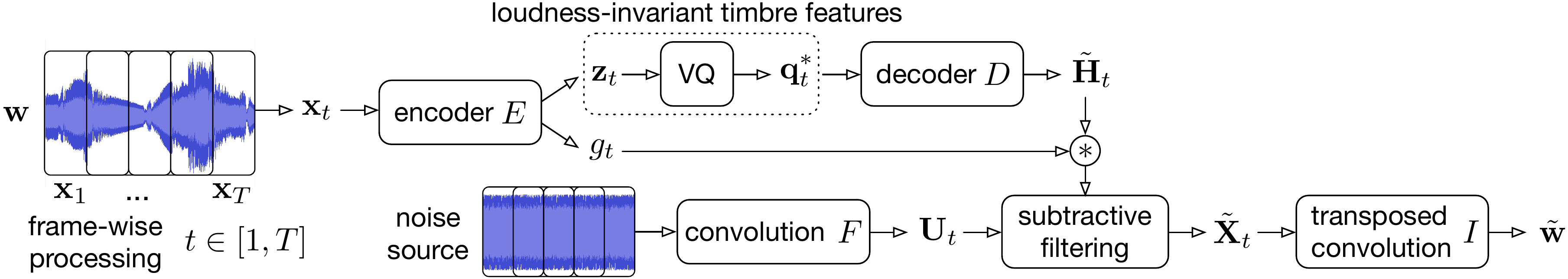}
\caption{Architecture of our proposed Vector-Quantized subtractive sound synthesis model.\label{fig:architecture}}
\end{figure*}

\subsection{Model Overview}\label{subsec:m_over}
We define an individual timbre through a corpus of audio files recorded for a target sound domain, for instance isolated or solo performances of a given instrument. A dataset of successive overlapping signal windows $\mathbf{x}_t\in\mathbb{R}^L$ is constructed by slicing input waveforms $\mathbf{w}$ of given duration into series $\{\mathbf{x}_1,\ldots,\mathbf{x}_T\}$. The encoder $E$ projects each of the $T$ windows of length $L$ into a continuous latent code $E(\mathbf{x}_t)=\mathbf{z}_t\in\mathbb{R}^{d_z}$, while reducing the dimensionality as $d_z<L$. A quantization estimator selects a vector $\mathbf{q}^*_t$ in the discrete embedding $\{\mathbf{q}^{_1},\ldots,\mathbf{q}^{_K}\}\in\mathbb{R}^{d_z*K}$ that is the closest match to $\mathbf{z}_t$. The decoder $D$ predicts filtering coefficients $D(\mathbf{q}^*_t) = \tilde{\mathbf{H}}_t\in\mathbb{R}^{N}$ that are applied to spectral frames $\mathbf{U}_t$ of a noise excitation, with $N$ the number of frequency bins. In order to disentangle loudness from the latent timbre embedding, the encoder predicts an additional scalar gain $g_t$. This architecture is depicted in \figref{fig:architecture} and the output time frames are filtered as
\begin{equation}
\tilde{\mathbf{X}}_t=g_t*\tilde{\mathbf{H}}_t . \mathbf{U}_t.
\label{eq:nsub_filt}
\end{equation}
The reconstruction is done by inversion of $\{\tilde{\mathbf{X}}_1,\ldots,\tilde{\mathbf{X}}_T\}$ into $\tilde{\mathbf{w}}$. This overlap-add uses the same stride as the encoder and the noise spectrum, and it can be performed for variable-length signals.

\subsection{Encoding Modules}\label{subsec:encoder}
The first layer of the encoder slices the input waveform into overlapping windows with a convolution of stride $S$ and Hanning kernel of size $L$ set as a power of 2. Every individual window is passed into a stack of downsampling convolutions with stride 2. One output layer predicts the latent features $\mathbf{z}_t$ and another infers the scalar gains $g_t$. The latent features are projected into the discrete embedding, yielding the quantization codes $\mathbf{q}^*_t$ sent to the decoder.

\subsection{Decoding Modules}\label{subsec:decoder}
Subtractive synthesis is performed by filtering an excitation with flat energy distribution. A uniform noise signal of the same length as $\mathbf{w}$ is converted into complex spectrum frames $\mathbf{U}_t$. We use a convolution $F$ with a stride $S$ and $N$ kernels of size $L$ corresponding to the Fourier basis. The first half of the bins is the real part and the other is the imaginary part. The series of quantized features $\mathbf{q}^*_t$ is processed by the decoder which predicts the series of filtering coefficients $\tilde{\mathbf{H}}_t$. The decoder is composed of an input stack of linear layers, a Recurrent Neural Network (RNN) and an output stack of linear layers. The predicted filters are scaled with the disentangled gains as $g_t * \tilde{\mathbf{H}}_t$ and applied to the noise spectrum. Synthesis from the filtered frames $\tilde{\mathbf{X}}_t$ is done by overlap-add. We use a transposed convolution $I$ of stride $S$ and $N$ kernels of size $L$ corresponding to the inverse Fourier basis. Such use of convolutional neural networks for time-frequency analysis and synthesis has previously been detailed for both music information retrieval\cite{nnaudio} and source separation\cite{fbank_sep} tasks.

\subsection{Model Objectives}\label{subsec:m_obj}
Our proposed model jointly optimizes waveform reconstruction and vector quantization with encoder commitment and codebook update losses. In order to evaluate the reconstruction, we use a multi-scale spectrogram loss over several STFT resolutions of magnitudes $l_n$ and the deep feature distance $d$ defined in Eqn \eqref{eq:percep_dist}. The different loss contributions are scaled by hyperparameters $\lambda_{0,1}$ for reconstruction terms and $\lambda_2$ for latent optimization, as
\begin{align}
\mathcal{L}=\lambda_{0}*\sum_n{\left\|l_n(\mathbf{w})-l_n(\tilde{\mathbf{w}})\right\|_1}+\lambda_{1}*d(\mathbf{w},\tilde{\mathbf{w}})\\ \nonumber
+\lambda_{2}*(\mathcal{L}_{codebook} + \beta*\mathcal{L}_{commit}).
\end{align}
\label{eq:nsub_obj}

\section{Experiments}\label{sec:exper}

\begin{table*}[!ht]
\begin{center}
\scalebox{0.9}{
\begin{tabular}{c|c;{1pt/1pt}c|c;{1pt/1pt}c|c;{1pt/1pt}c|c;{1pt/1pt}c} 
\hline
scores                          & \multicolumn{2}{c|}{classification accuracy} & \multicolumn{2}{c|}{DTW~$f_0$} & \multicolumn{2}{c|}{DTW loudness} & \multicolumn{2}{c}{LSD}  \\ 
\hline
\textbf{targets}~$|$~models     & baseline        & VQ-VAE          & baseline & VQ-VAE                       & baseline        & VQ-VAE       & baseline & VQ-VAE                 \\ 
\hline
\textbf{bassoon}   &  0.4256 &  \textbf{0.6456} & 4.8e-4 & \textbf{4.7e-4} & \textbf{2.5e-4} & \textbf{2.5e-4} & \textbf{0.4193} & 0.4387 \\ 
\hline
\textbf{cello}      & 0.3182  & \textbf{0.5896} & \textbf{4.7e-4} & 5.5e-4 & 2.2e-4 & \textbf{1.9e-4} & \textbf{0.4500} & 0.4853 \\ 
\hline
\textbf{clarinet}   & 0.3962 & \textbf{0.6811} & \textbf{4.1e-4} & 4.6e-4 & 3.8e-4 & \textbf{2.4e-4} & 0.4341 & \textbf{0.4303}       \\ 
\hline
\textbf{double-bass} & 0.1190  & \textbf{0.4298}  & 6.6e-4  & \textbf{6.0e-4}  & \textbf{2.5e-4}  & 3.0e-4  & \textbf{0.4313}  & 0.4346  \\ 
\hline
\textbf{flute} & 0.4999  & \textbf{0.6765}  & \textbf{6.5e-4}  & 8.8e-4  & \textbf{2.5e-4}  & 2.7e-4  &  \textbf{0.3623} & 0.3735  \\ 
\hline
\textbf{horn} & 0.4104  & \textbf{0.5861}  & \textbf{4.4e-4}  & 5.2e-4  & 2.4e-4  & \textbf{1.8e-4}  & \textbf{0.3910}  & 0.4409  \\ 
\hline
\textbf{oboe} & 0.6610  & \textbf{0.7441}  & \textbf{6.3e-4}  & 6.5e-4  & \textbf{3.1e-4}  & 3.3e-4  & \textbf{0.3679}  &  0.3840 \\ 
\hline
\textbf{trumpet}   & 0.5126  & \textbf{0.6041} & \textbf{3.7e-4} & 4.0e-4 & 5.8e-4 & \textbf{3.7e-4} & \textbf{0.3625} & 0.3719   \\ 
\hline
\textbf{viola} & \textbf{0.6409}  & 0.5689  & \textbf{4.1e-4}  & 4.2e-4  & 2.4e-4  & \textbf{2.3e-4}  & 0.4606  & \textbf{0.3946}  \\
\hline
\textbf{violin}    & 0.7434   & \textbf{0.7960} & \textbf{3.7e-4} & 4.5e-4 & 4.1e-4 & \textbf{2.8e-4} & 0.5546 & \textbf{0.5500}  \\ 
\hline
\textit{instrument average} & \textit{0.4727}  & \textit{\textbf{0.6321}}  & \textit{\textbf{4.8e-4}}  & \textit{5.4e-4}  & \textit{3.1e-4}  & \textit{\textbf{2.6e-4}}  & \textit{\textbf{0.4233}}  & \textit{0.4303}  \\
\hline \hline
\textbf{singing}    & N.A.  & N.A. & \textbf{3.2e-4} & 3.6e-4 & \textbf{2.7e-4} & 2.8e-4 &       \textbf{0.5477} &  0.5523 \\
\hline
\end{tabular}}
\end{center}
\caption{Score comparison of the VQ-VAE model against the baseline auto-encoder.  Classification accuracy assesses the transfer to the instrument target of each model. DTW measures the distance between the source and audio transfer curves of $f_0$ and loudness. LSD evaluates the test set reconstruction error in the target domain. Bold denotes the best score.}
\label{tab:bench}
\end{table*}

\subsection{Datasets}\label{subsec:dataset}
In order to learn the individual timbre of instruments, we rely on multitrack recordings of music performances from two datasets, namely URMP\cite{urmp} and Phenicx\cite{phedb}. They both provide isolated audio for \textit{bassoon}, \textit{cello}, \textit{clarinet}, \textit{double-bass}, \textit{flute}, \textit{horn}, \textit{oboe}, \textit{trumpet}, \textit{viola} and \textit{violin}.

To learn the singing voice timbre representation, we use the recordings from the VocalSet\cite{vocalset} database which provides 9 female and 11 male singers individually performing several techniques and pitches. We discard the noisiest techniques \textit{breathy}, \textit{inhaled}, \textit{lip-trill}, \textit{trillo}, \textit{vocal fry} and merge all others in the same timbre domain.

To experiment with voice-controlled sound synthesis, we use some examples of the VocalSketch\cite{vocalsketch} database, which were given as source inputs to models pretrained on instruments. Vocal imitations were not used as training data, but as crowd-sourced examples of untrained human voices expressing some diverse sound concepts.

\subsection{Perceptual Audio Loss}\label{subsec:percep_loss}
Using the dataset of just-noticeable audio differences and human ratings \cite{perceploss}, we re-implemented the deep feature distance in \textit{PyTorch} (codes and pretrained parameters of $d$ are provided\footnote{\url{https://github.com/adrienchaton/PerceptualAudio_Pytorch}}). To use this loss as a reconstruction objective for music performances recorded at various volumes, we apply a random gain to the training audio pairs so that the learned distance is invariant to audio levels. As this criterion was trained for several perturbations including additive noises and reverberation, the model optimizes additional acoustic cues to generate audio signals that are consistent with the training dataset. We observe that vocal imitations recorded in uncontrolled conditions can be transferred into musical sounds which do not exhibit the input noise found in VocalSketch.

\subsection{Training Details}\label{subsec:training_det}
All audio examples are first downsampled to 22kHz in mono format. The subsets corresponding to each individual timbre, either instrumental or singing voice, are split into training and test data (15\%). We remove silences and concatenate the trimmed audio. Segments $\mathbf{w}$ of 1.5 seconds are randomly sampled in the training data and collated into mini-batches of size 20 for training the VQ-VAE. We optimize the model for 150,000 iterations with the ADAM optimizer and a learning rate of 2e-4.

The model is defined with window size $L=2048$, stride $S=512$ and $N=L+2$ which corresponds to the real and imaginary parts of the halved complex spectrum. The encoder has 7 downsampling convolutions of stride 2, with increasing output channel dimension from 32 to 256 and kernel size 13. One output layer maps to latent features of size $d_z=128$ and another pair of linear layers outputs the scalar gains. The vector quantization space is a codebook of size $K=1024$. The decoder has two blocks of 4 linear layers with a constant hidden dimension of 768 that are interleaved with intermediate Gated Recurrent Units of the same feature size. The output of the decoder is a linear layer that produces $N$ filtering coefficients which are passed into a \textit{sigmoid} activation and \textit{log1p} compression. The convolutions $F$ and $I$ are initialized as the linear STFT and its inverse, future experiments could include using different frequency scales or training their kernels. The multi-scale sprectrogram reconstruction is computed for STFTs with a hop ratio of 0.25 and window sizes of [128, 256, 512, 1024, 2048]. We adjust the $\lambda$ strengths in order to balance the initial gradient magnitudes of each objective, accordingly $\lambda_0=\lambda_2=1$ and $\lambda_1=0.2$. The latent loss uses an encoder commitment strength of $\beta=0.25$.

\subsection{Classifier Model}\label{subsec:classif}
In order to evaluate the timbre transfer task, we train a reference classifier on the 10 target instruments. We adapt the baseline proposed in \cite{vocalset} to perform short-term predictions rather than predicting a single label per file. Our classifier predicts a label every non-overlapping frame of 4096 samples which amounts to a context of about 185ms. The model was trained with pitch-shifting data augmentation and achieves 85\% test set frame-level accuracy at predicting the correct instrument label.

\subsection{Evaluation}\label{subsec:eval}
The performance of our VQ-VAE is quantitatively compared against a \textit{baseline} deterministic auto-encoder without vector quantization. Since its latent space is continuous, the disentangled gain prediction did not improve the baseline and is as well removed. Besides that, it shares the same encoder and decoder architectures and only optimizes reconstruction costs. We compare the models in terms of spectrogram reconstruction quality in the learned timbre domain and transfer quality from other sources.

\section{Results}\label{sec:results}

\subsection{Comparative Model Evaluation}\label{subsec:compare_base}
The test set reconstruction quality of the models is evaluated by comparing the spectrogram magnitudes of the input and output waveforms using the Log-Spectral Distance (LSD). The instrument timbre transfer accuracy is evaluated by auto-encoding every other instrument subsets from URMP and Phenicx (besides the trained target) and every singing excerpts from VocalSet and predicting the instrument label of the synthesized audio with the pretrained reference classifier. The accuracy is reported with respect to the target instrument, and aims to be maximized. In addition, the source $f_0$ and loudness curves are compared with those of the audio transfer. We use the Dynamic Time Warping (DTW) distance to measure how well the model preserves pitch and loudness independently from transferring timbre. The DTW score is normalized across audio excerpts by scaling the time series in unit range and averaging by the lengths of the DTW paths. For the model trained on singing voice, we transfer audio from all the instrument subsets and only report the average DTW distances.

As detailed in \tabref{tab:bench}, the discrete representation of the VQ-VAE consistently improves the unsupervised timbre transfer accuracy in comparison with the baseline auto-encoder. For inference on other source domains, our proposed model solely uses a fixed basis of latent features learned from the spectral distribution of a given timbre. As a result, the quantization enforces audio transfer of the target timbre properties. We also observe that the disentangled gain prediction tends to improve the reconstruction of loudness, as shown by a lower average DTW distance for the VQ-VAE model. However, we did not constrain the model to rely on an explicit estimate of the fundamental frequency. Since it is not disentangled from the representation, we observe that quantization comes at the expense of a lesser accurate reconstruction of the pitch than for the continuous baseline model. Notably, in the VQ-VAE this property is bound to the trained instrument tessitura. The overall reconstruction quality in the target timbre, assessed with the test set LSD, is similar for both auto-encoders.

Besides the quantitative evaluation of the discrete representation against the baseline auto-encoder, we note two additional benefits of feature quantization. When processing out-of-domain audio of lower quality, such as vocal imitations recorded in uncontrolled conditions, the transfer ability is paired with denoising. Indeed, acoustically inconsistent features are discarded in the latent projection to a trained domain such as musical studio recordings. This facilitates the use of timbre transfer from diverse recording environments such as for voice controlled synthesis. Moreover, we show that learning a discrete latent representation enables a direct mapping to acoustic descriptors as an other mean of high-level synthesis control.

\begin{figure}[ht]
\centering
\includegraphics[width=1.\columnwidth]{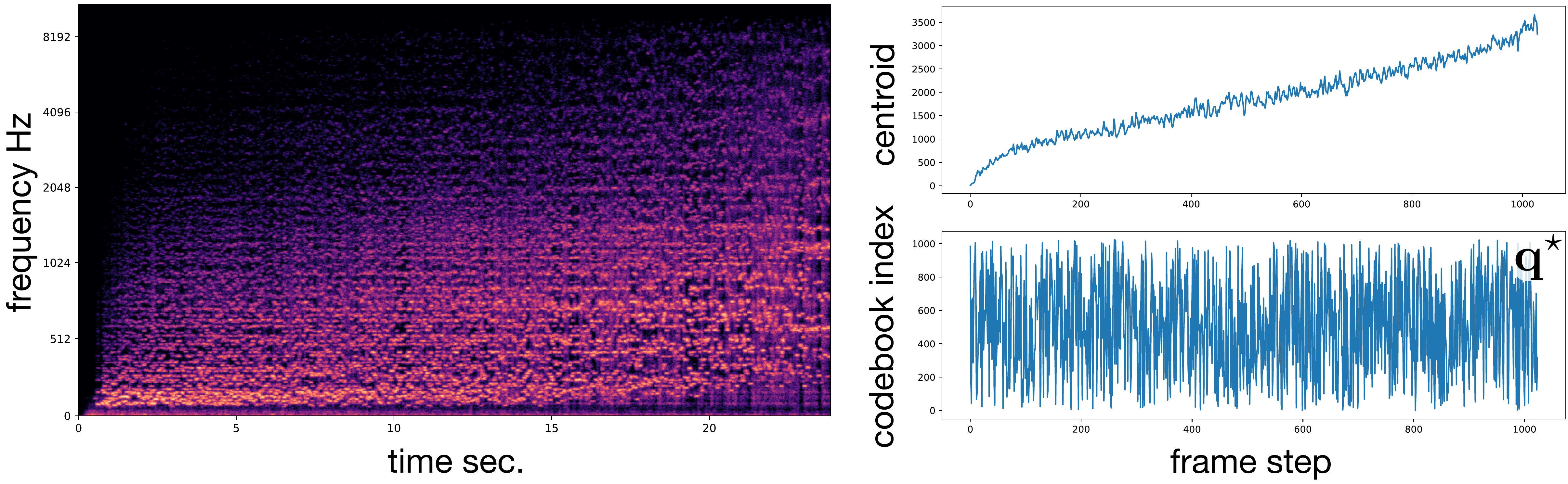}
\caption{The spectrogram and centroid of an audio synthesized with an increasing centroid target in the violin representation. The corresponding series of embedding indexes $\mathbf{q}^\star$ does not exhibit any structure but can be arranged consistently with the acoustic descriptor target.\label{fig:centroid}}
\end{figure}

\subsection{Descriptor-Based Timbre Synthesis}\label{subsec:VQ_audio}
In comparison with the baseline auto-encoder, the VQ-VAE decoder optimizes generation solely based on a discrete latent codebook. We introduce a mapping method for controllable sound synthesis (detailed in the supplementary material\footnote{\url{https://adrienchaton.github.io/VQ-VAE-timbre/}\label{url_git}}). Each embedding vector $\mathbf{q}^{_k}$ with $k\in[1,K]$ approximately corresponds to a short-term timbre feature and a spectral filter $\tilde{\mathbf{H}}^{k}=D(\mathbf{q}^{_k})$. Given that the decoder has a RNN, some temporal relationships are introduced in the overlap-add subtractive synthesis. We decode a series of an individual feature $\{\mathbf{q}^{_k},\ldots,\mathbf{q}^{_k}\}$ and compute the average acoustic descriptor value $\tilde{a}^{_k}$. After analyzing every latent vector, we obtain the mapping $\{\mathbf{q}^{_1},\ldots,\mathbf{q}^{_K}\} \leftrightarrow \{\tilde{a}^{_1},\ldots,\tilde{a}^{_K}\}$.

We can perform acoustic descriptor-based synthesis from a target $a_i$ of any length $M$ with $i\in[1,M]$ by selecting the nearest values in the discrete mapping $\tilde{a}^\star_i$ and decoding the corresponding series of latent features $\mathbf{q}^\star_i$. The mark $^\star$ is used here to denote the nearest embedding elements to the descriptor target, whereas in Eqn \eqref{eq:near_code} the selection of $\mathbf{q}^*$ is done by matching with the encoder output. Using such mapping, we show in \figref{fig:centroid} that we can control a VQ-VAE model of violin with an increasing \textit{centroid} target. The decoded audio has a consistent spectrogram and synthesized centroid. We also observe that the acoustically ordered series of latent features corresponds to an unordered traversal of the discrete embedding. In other words, the index positions in the quantization space do not correlate to acoustic similarities, which are only provided by our proposed mapping method.

\begin{figure}[ht]
\centering
\includegraphics[width=1.\columnwidth]{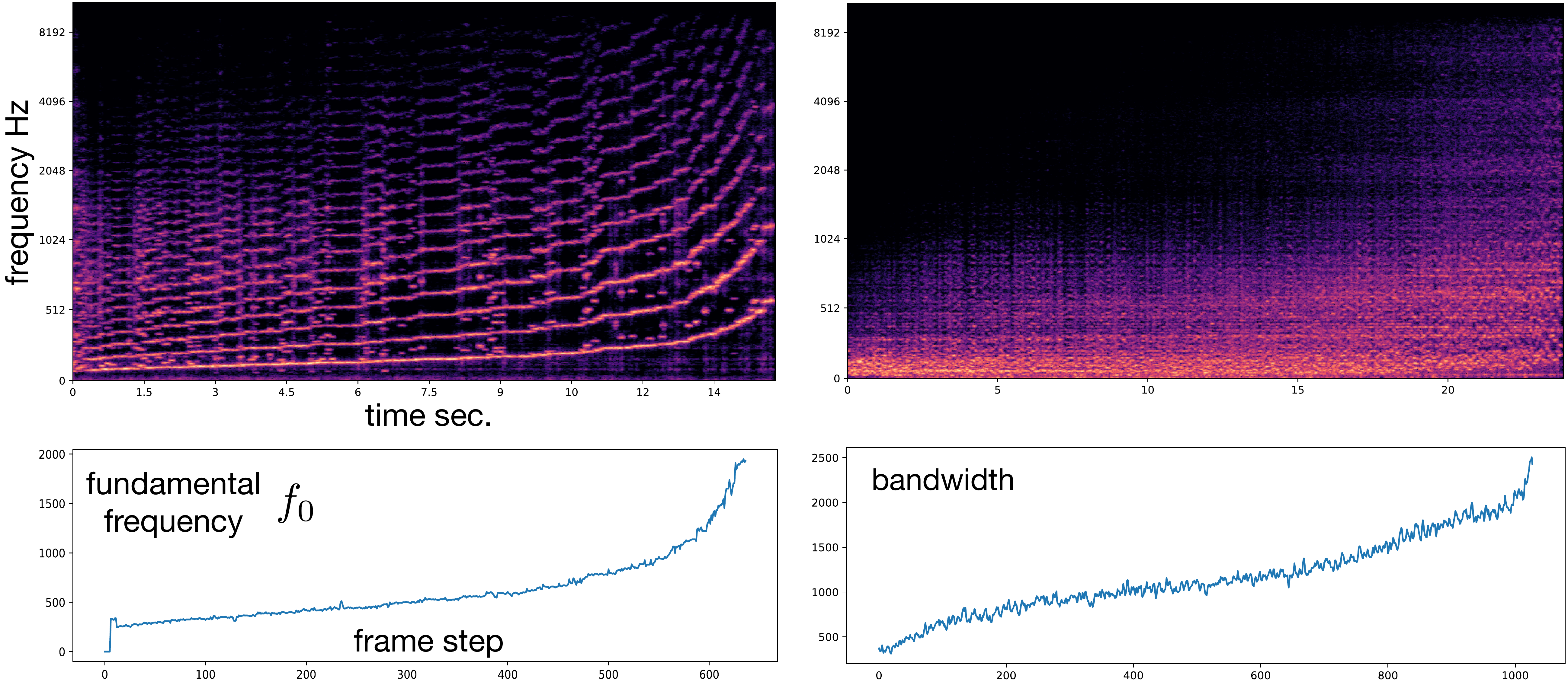}
\caption{Controlling the sound synthesis of the violin by the $f_0$ (left) and the cello by the bandwidth (right).\label{fig:vn_vc}}
\end{figure}

This analysis can be performed for other acoustic descriptors and other instrument representations. In \figref{fig:vn_vc}, we depict the control of the VQ-VAE model by a target defined either with \textit{fundamental frequency} for the violin or with \textit{bandwidth} for the cello. Our proposed model does not rely on $f_0$ conditioning in order to process diverse audio sources, such as vocal imitations without pitch. However, we show that the fundamental frequency can be controlled by mapping the unsupervised representation. Our proposed method yields an approximate decomposition of the acoustic properties of an individual timbre, it allows high-level and direct controls for sound synthesis.

\section{Conclusion}\label{sec:conclusion}

We have introduced a raw waveform auto-encoder to learn a discrete representation of an individual timbre that is disentangled from loudness. It can be used for unsupervised transfer of musical instrument performances and singing voice. The model generates audio by subtractive sound synthesis, a technique which neither restricts the types of signals nor the duration that can be processed. The spectral distribution of a timbre is quantized with a set of short-term latent features that are decoded into noise filtering coefficients. This discrete representation can be mapped to acoustic properties in order to perform direct descriptor-based synthesis. Some descriptor targets can be matched with latent features that are decoded into signals with the desired auditory qualities. For instance, the unsupervised model can be controlled with the fundamental frequency. In addition, we experiment with transferring vocal imitations into an instrument timbre as an example of voice-controlled sound synthesis. Audio samples are provided in the supporting GitHub page\textsuperscript{\ref{url_git}}.

\section{Acknowledgements}\label{sec:acknow}

This work was done under a Japanese Society for Promotion of Science (JSPS) short-term fellowship. We thank the JSPS and The University of Tokyo for their outstanding support.

\bibliography{main}

\end{document}